\documentclass[10pt,prd,aps,superscriptaddress,floatfix,nofootinbib,eqsecnum]{revtex4-2}

\pdfoutput=1

\usepackage{multirow}
\usepackage{amsmath}
\usepackage{amsfonts}
\usepackage{amsmath}
\usepackage{amssymb}
\usepackage{bm}
\usepackage{float}
\usepackage{adjustbox}
\usepackage{dcolumn}
\usepackage{graphicx}
\usepackage[utf8]{inputenc}
\usepackage{latexsym}
\usepackage{rotating}
\usepackage{hyperref}
\usepackage{subfigure}
\usepackage{color}
\usepackage{changes}
\usepackage{comment}
\usepackage{verbatim}
\usepackage{comment}
\usepackage{empheq}
\usepackage{csquotes}
\usepackage{physics}
\usepackage{soul}
\usepackage{amsmath,latexsym}
\usepackage{mathrsfs}
\usepackage{orcidlink}

\usepackage{booktabs}  

\begin{document}

\title{Comparing Mass-Varying Neutrino Dark Energy Models with DESI DR2: Cosmological Constraints and Bayesian Evidence}

\author{Hemanshi Bundeliya} \email {hemanshibundeliya07@gmail.com}\affiliation{Department of Physics, Lovely Professional University,  Phagwara, Punjab, 144411, India}

\author{Gaurav Bhandari}\email{bhandarigaurav1408@gmail.com}\affiliation{Department of Physics, Lovely Professional University,  Phagwara, Punjab, 144411, India}

\author{Mohammad Yarahmadi}\email{yarahmadi.mhd@u.ac.ir}
\affiliation{ Department of Physics, Lorestan University, Iran}

\author{Mahdi Sepahvand}\email{mahdisepahvand7098@gmail.com}
\affiliation{ Department of Physics, Lorestan University, Iran}

\author{V. K. Sharma}\email{vipin.33912@lpu.co.in}
\affiliation{Department of Physics, Lovely Professional University, Phagwara, Punjab, 144411, India}
\affiliation{ International Center for High Energy Physics and Applications, Lovely Professional University, Phagwara, Punjab, 144411, India}
\affiliation{Research Center of Astrophysics and Cosmology, Khazar University, Baku, AZ1096, 41 Mehseti Street, Azerbaijan}

\author{S. D. Pathak}\email{sdpathak@lko.amity.edu}\affiliation{Amity School of Applied Sciences,\\ Amity University Uttar Pradesh, Lucknow Campus, Lucknow, 226028, India}

\begin{abstract}
We investigate late-time cosmic acceleration in mass-varying neutrino (MaVaN) dark energy scenarios, where the neutrino mass is coupled to a dynamical quintessence field through a conformal interaction. Starting from the coupled conservation equations, we derive the corresponding background evolution equations for three related scenarios: a constant dark-energy equation of state with a constant coupling (Case-I), an exponential neutrino mass linked to an exponential potential under an adiabatic minimum-tracking approximation (Case-II), and a CPL-type time-varying equation of state with constant coupling (Case-III). For each case we obtain the neutrino and dark energy density evolution and the corresponding Hubble parameter, and use an MCMC analysis (emcee) to test the models against DESI DR2 BAO data, the compressed Planck CMB likelihood, and three supernova compilations (Pantheon+, DES-SN5YR, Union3), taken one at a time. Across the three cases, $H_0$ and $\Omega_{m0}$ move together in a way set mainly by which supernova sample is used, following the pattern already seen in other DESI DR2 analyses. A nonzero coupling stays consistent with the data in most fits, though its size, and sometimes its sign, depend on how it enters the model and on the priors chosen for the underlying parameters. The resulting neutrino mass bounds range from about $0.06$ to $0.3\,\mathrm{eV}$, depending on the case and dataset combination. Comparing the Bayesian evidence of each case with flat $\Lambda$CDM gives a mixed picture, with no case preferred throughout, so we regard these results as suggestive rather than firm evidence for a coupling between neutrinos and dark energy.
\end{abstract}
\maketitle
\section{Introduction}

Over the last few decades,our understanding of the universe has evolved dramatically. What was once based mainly on theoretical ideas and limited observations has now become a precise and data driven science \cite{10.1093/oso/9780198526827.001.0001}. Advanced observations from the Cosmic Microwave Background (CMB), large-scale galaxy surveys, gravitational lensing, Type-la supernovae, and baryon acoustic oscillations (BAO) have helped shape a detailed picture of the cosmos \cite{aghanim2020planck,Spergel_2003,Perlmutter_1999,Eisenstein_2005}. These combined results support what is known as the $\Lambda$CDM model, suggesting that the universe is nearly flat and mostly made up of components that we cannot see directly. According to this model, only about $5\%$ of the universe consists of ordinary matter like stars and planets. Roughly $25\%$ is dark matter,which interacts through gravity but does not emit light. The remaining $\sim 70\%$ is dark energy-a mysterious form of energy believed to be driving the accelerated expansion of the universe \cite{Percival_2010,Abbott_2022,Tegmark_2004,Freedman_2003,Weinberg_2013}.

Although the $\Lambda$CDM model matches observations remarkably well, it leaves important questions  \cite{Perivolaropoulos2022,efstathiou2024challengeslambdacdmcosmology}. The simplest explanation for dark energy is the cosmological constant $\Lambda$, interpreted as a constant vacuum energy that fills space. However, this idea raises deep theoretical issues \cite{Martin:2012bt,SolaPeracaula:2022hpd,Bernardo:2022cck}. One of them, the fine-tuning problem, highlights the enormous gap between the observed value of $\Lambda$ and the predictions from quantum field theory. Another, the coincidence problem, asks why dark energy became dominant only in the relatively recent cosmic past \cite{Weinberg:1988cp}. These challenges have encouraged us to explore more dynamic explanations of dark energy — especially models in which dark energy interacts with other fields or particles \cite{COPELAND_2006,Peebles_2003,Capozziello:2025qmh,Chaudhary:2025pcc,Chaudhary:2025bfs,KumarSharma:2026ppi}.

Among these possibilities, scenarios where dark energy interacts with neutrinos have gained considerable attention \cite{Fardon:2003eh,Brookfield_2006}. Neutrinos were produced in abundance in the early universe, interact extremely weakly, and transition from relativistic to non-relativistic states as the universe expands \cite{KolbTurner1990,Dodelson2003,LesgourguesPastor2006}. Importantly, experiments have confirmed that neutrinos have a non-zero mass through oscillation measurements \cite{SuperK1998,SNO2002}, although these masses are extremely small and their fundamental origin remains unknown \cite{PDG2022,MohapatraPal2004,GiuntiKim2007}. In Mass Varying Neutrino (MaVaN) models, Neutrino mass is neither fixed but evolves over time through its coupling with a scalar field, Which also plays the role of dark energy. When neutrinos slow down and become non-relativistic, their changing mass begins to influence the scalar field, shaping its evolution \cite{Peccei2005}.

This coupling can cause the scalar field to settle into a stable configuration at late times,effectively acting like dark energy. Interestingly, this mechanism naturally ties the Onset of cosmic accelerations to the moment when neutrino becomes non-relativistic, Potentially offering a solution to the coincidence problem. In this way, MaVaN models create a bridge between particle physics and cosmology, suggesting that the key to understanding dark energy may be hidden in the physics of neutrinos \cite{Wang2016}.

Recent observational advances provide a powerful opportunity to test such scenarios. The Dark Energy Spectroscopic Instrument (DESI) is mapping the three-dimensional distribution of galaxies and quasars with unprecedented precision over a wide redshift range. Its measurements of BAO and large-scale structure place stringent constraints on the cosmic expansion history and the growth of structure \cite{DESI:2025zgx,Chaudhary:2025bfs,Chaudhary:2025pcc}. Since MaVaN models predict deviations from $\Lambda$CDM in both the background expansion and potentially in perturbations, the latest DESI Data Release 2 (DR2) results offer a timely dataset for confronting these models with observations.

Several theoretical realizations of the Mass-Varying Neutrino (MaVaN) scenario have been proposed in the literature. In the original formulation, it was shown that a scalar field coupled to neutrinos can dynamically generate an effective potential whose minimum evolves with the neutrino energy density, thereby linking the onset of cosmic acceleration to neutrino physics \cite{Fardon2004}. As the Universe expands and neutrinos transition from relativistic to non-relativistic behavior, the coupling between the scalar field and the neutrino sector becomes dynamically relevant, which can naturally trigger late-time acceleration. Subsequent studies investigated the cosmological implications of such couplings in greater detail, including the effects on background evolution and perturbations \cite{Brookfield2006,Amendola2008}. However, it has also been pointed out that certain realizations of MaVaN models may suffer from dynamical instabilities, such as the formation of neutrino density perturbations or ``neutrino lumps,'' depending on the strength of the coupling and the form of the scalar potential \cite{Afshordi2005}.

From the observational perspective, the viability of MaVaN scenarios must be tested against increasingly precise measurements of the cosmic expansion history. Over the past two decades, large galaxy surveys and distance measurements have significantly improved our ability to constrain cosmological models. In particular, Baryon Acoustic Oscillation (BAO) measurements provide one of the most robust probes of the late-time expansion history because they offer a standard ruler that is relatively insensitive to astrophysical systematics. 

The Dark Energy Spectroscopic Instrument (DESI) represents a major step forward in this direction. By mapping the large-scale distribution of millions of galaxies and quasars across a wide redshift range, DESI provides high-precision measurements of BAO and large-scale structure, enabling stringent tests of dark energy models. The recent Data Release 2 (DR2) extends these measurements using multiple tracers, including luminous red galaxies, emission line galaxies, quasars, and the Lyman-$\alpha$ forest, thereby offering a powerful dataset to constrain models that modify the late-time expansion history \cite{karim2025desi}.

Complementary constraints are provided by Type Ia supernova observations, which serve as standardizable candles for measuring cosmic distances. Modern compilations such as the Pantheon$^{+}$ sample \cite{brout2022pantheon}, the five-year Dark Energy Survey supernova dataset \cite{abbott2024dark}, and the Union compilation \cite{rubin2025union} contain thousands of supernova observations spanning a broad redshift range. When combined with Cosmic Microwave Background constraints on the early Universe, these datasets provide a powerful framework for testing dynamical dark energy models and probing possible deviations from the standard $\Lambda$CDM scenario.

In this work, we investigate late-time cosmic acceleration within the framework of mass-varying neutrino (MaVaN) cosmologies, where the neutrino sector interacts with a dynamical dark-energy scalar field. Starting from the coupled conservation equations, we derive the cosmological evolution for constant and time-dependent dark energy equation of state (EoS) as well as for an exponential neutrino-mass model coupled to an exponential potential. For the latter scenario, we employ the adiabatic minimum-tracking approximation to obtain analytical expressions for the neutrino and dark-energy densities and the corresponding expansion history. To place observational constraints on the model, we perform a Markov Chain Monte Carlo (MCMC) analysis using DESI DR2 BAO measurements.

The paper is organized as follows. In Sec.~\ref{a}, we present the theoretical framework of mass-varying neutrino cosmologies and the coupling between the neutrino and scalar-field sectors. In Sec. \ref{b}, we derive the cosmological evolution equations and investigate three representative scenarios: constant dark-energy equation of state with constant interaction strength, exponential mass-varying neutrinos coupled to an exponential potential, and a time-dependent dark energy EoS described by the CPL parametrization. For each case, we obtain the corresponding analytical expressions for the neutrino and dark energy densities, the dark energy evolution function, and the normalized Hubble parameter. In Sec.~\ref{DATASET}, we describe the observational datasets, statistical methodology, and parameter estimation procedure employed in our analysis. Finally, we summarize our main results and conclusions in Sec.~\ref{sec_5}.

We prefer to work with (-  +  +  +) metric signature, and the natural units $c=\hbar=1$.

\section{Theoretical framework and field equations }\label{a}
We consider a conformally coupled scalar field interacting with matter through the action in
Einstein frame 
\begin{equation}
\label{actionneu}
S = \int d^4x \, \sqrt{-g} 
\left[ 
\frac{1}{2} M_P^2 R 
- \frac{1}{2} g^{\mu\nu} \partial_\mu \phi \, \partial_\nu \phi 
- V(\phi)
\right] 
+ \sum_j S_j \big[ B_j^2(\phi) g_{\mu\nu}, \psi_j \big],
\end{equation}
where $R$ is the usual Ricci scalar, 
and $M_P = (8\pi G)^{-1/2} = 2.4 \times 10^{18}\,\mathrm{GeV}$ is the 
reduced Planck mass. The scalar field $\phi(t)$ represents a homogeneous quintessence field 
with potential $V(\phi)$, while $\psi_j$ denotes the various matter field species. The interaction between the scalar field and the matter sector is 
introduced through a conformal rescaling of the metric in each matter action, $\tilde{g}_{\mu\nu}^{(j)} = B_j^2(\phi)\, g_{\mu\nu}$, where $B_j(\phi) > 0$ is the conformal coupling function. In this framework, the gravitational sector remains minimally coupled, whereas 
matter fields experience a $\phi$-dependent metric.

For the maximally symmetric flat background spacetime, the Friedmann–Lemaître–Robertson–Walker (FLRW) metric is represented as,
\begin{equation}
\label{a242}
ds^{2} = - dt^{2} + a^{2}(t)\,\delta_{ij}\, dx^{i} dx^{j},
\end{equation}
where $a(t)$ is time dependent scale factor and the speed of light $c=1$.

Using the variation of action \eqref{actionneu} with respect to the metric $g_{\mu\nu}$ and the scalar field $\phi$ provide the evolution equations, namely Friedmann, Raychaudhari and Klein-Gordon equation, respectively
\begin{align}
&H^{2}
=
\frac{1}{3M_{P}^{2}}
\left(
\frac{1}{2}\dot{\phi}^{2} + V + \sum_{j}\rho^{(j)}
\right), \label{freid}\\
&\dot{H}
=
-\frac{1}{2M_{P}^{2}}
\left(
\dot{\phi}^{2} + \sum_{j}\left(\rho^{(j)} + P^{(j)}\right)
\right),\label{ray}\\
&\ddot{\phi} + 3H\dot{\phi} + V_{,\phi}
=
-\frac{B_{,\phi}}{B}\left(\rho^{(i)} - 3P^{(i)}\right).\label{klein}
\end{align}
where overdots represent the derivative w.r.t time and $\rho^{(i)}$ and $P^{(i)}$ are the energy density and pressure of the interacting component, respectively, and $V_{,\phi} = dV/d\phi$.

The energy conservation equation for the combined fluid of dark energy and other species is given as
\begin{equation}
\dot{\rho}_{total} + 3H\left(\rho_{total} + P_{total}\right) = 0,
\end{equation}
and for the individual component the continuty equation is written as  
\begin{equation}\label{continuity}
\dot{\rho}^{(i)} + 3H\left(\rho^{(i)} + P^{(i)}\right)
= \frac{B_{,\phi}}{B}\dot{\phi}\left(\rho^{(i)} - 3P^{(i)}\right).
\end{equation}
It is interesting to consider scalar field-neutrino interaction from the given action in Eq.(\ref{actionneu}) as coupled scalar field model studied in \cite{amendola2000, Farrar_2004,wetterich1994cosmonmodelasymptoticallyvanishing}.

Considering the neutrino sector, the neutrino mass is assumed to be field-dependent, \( m_\nu = m_\nu(\phi) \). In a cosmological setting, neutrinos are described within a kinetic framework by a phase-space distribution function \( f(x^i, p^i, t) \), which satisfies the collisionless Boltzmann (Liouville) equation~\cite{Dodelson,Weinberg}. For a homogeneous and isotropic background, the distribution function depends only on the magnitude of the comoving momentum, \( q = a p \). Under these assumptions, the Boltzmann equation yields the neutrino energy density and pressure as \cite{Mukhanov,Liddle}

\begin{equation}\label{neurho}
\rho_\nu = \frac{1}{a^4} \int q^2 \, dq \, d\Omega \, \mathcal{E}(q) \, f_0(q),
\qquad
p_\nu = \frac{1}{3a^4} \int q^2 \, dq \, d\Omega \, f_0(q) \, \frac{q^2}{\mathcal{E}(q)},
\end{equation}
where $f_0(q)$ denotes the background Fermi–Dirac distribution and the single-particle energy satisfies, $
\mathcal{E}^2 = q^2 + a^2 m_\nu^2(\phi)$.
These expressions provide the general kinetic definitions of $\rho_\nu$ and $p_\nu$, explicitly incorporating the mass variation of neutrino through $m_\nu(\phi)$.

Considering isotropy ($\int d\Omega = 4\pi$) and introducing the dimensionless variables
\begin{equation}
x = \frac{\mathcal{E}}{T_\nu}, 
\qquad 
\zeta = \frac{m_\nu(\phi)}{T_\nu},
\end{equation}
with $T_\nu(a) = T_{\nu,0}/a$, the neutrino energy density and pressure can be written compactly as
\begin{equation}
\rho_\nu(a) = \frac{2}{\pi^2} T_\nu^4(a)\, I_\varepsilon(\zeta),
\qquad
p_\nu(a) = \frac{2}{3\pi^2} T_\nu^4(a)\, I_{3/2}(\zeta).
\end{equation}
The dimensionless Fermi–Dirac integrals are defined as
\begin{equation}
I_\varepsilon(x) =
\int_{\zeta}^{\infty}
\frac{x^2 \sqrt{x^2 - \zeta^2}}{e^x + 1}\, dx,
\qquad
I_{3/2}(\zeta) =
\int_{\zeta}^{\infty}
\frac{(x^2 - \zeta^2)^{3/2}}{e^x + 1}\, dx.
\end{equation}
Although these integrals do not admit closed-form expressions for arbitrary $\zeta$, their limiting behaviours can be obtained analytically in the relativistic ($\zeta \ll 1$) and non-relativistic ($\zeta \gtrsim 1$) regimes. In our recent research, we have discussed the  relativistic regime during inflationary era \cite{CPC:10.1088/1674-1137/ae432b}. 

From the equations \eqref{neurho}, the continuity equation gives,
\begin{equation} \label{2.11}
\dot{\rho}_\nu + 3H\left(\rho_\nu + P_\nu\right)
=
\frac{d \ln{m_\nu(\phi)}}{d \phi}\dot{\phi}\left(\rho_\nu - 3P_\nu\right).
\end{equation}

Thus \eqref{continuity} is the same as the mass varying neutrino continuity equation provided that $B(\phi)=m_\nu(\phi)/M$, where $M$ is a scaling factor of mass. Under this identification, the scalar field couples directly to the neutrino sector through the field dependence of the neutrino mass. As a consequence, the Klein--Gordon equation \eqref{klein} is modified and can be written in terms of an effective potential as,
\begin{equation}\label{equation of motion}
\ddot{\phi} + 3H\dot{\phi} + V_{\mathrm{eff},\phi} = 0,
\end{equation}
where the scalar dynamics are governed not only by the bare potential $V(\phi)$ but also by the neutrino contribution. Therefore, the effective potential of \eqref{equation of motion} will become,
\begin{equation}\label{2.13}
    V_{\mathrm{eff},\phi}
=
V_{,\phi}
+
\frac{m_{\nu,\phi}(\phi)}{m_\nu(\phi)}
\left(\rho_\nu - 3P_\nu\right).
\end{equation}
Here, the quantity $\rho_\nu - 3p_\nu$ corresponds to the trace of the neutrino energy-momentum tensor, which controls the strength of the coupling between the scalar field and neutrinos based on non-relativistic scenarios. We will discuss this next.

\section{Exchanges of energy between scalar field and Mass Varying Neutrino}\label{b}


In the early Universe, neutrinos can be treated as relativistic particles, satisfying \( m_\nu \ll T_\nu \). In this regime, their equation of state approaches, \( P_\nu = \rho_\nu/3 \). Consequently, the trace of the neutrino energy-momentum tensor vanishes,
\begin{equation}
\rho_\nu - 3P_\nu \approx 0,
\end{equation}
which implies that the coupling between the neutrino sector and the scalar field becomes negligible. As a result, the Klein–Gordon equation reduces to its standard uncoupled form,
\begin{equation}
\ddot{\phi} + 3H\dot{\phi} + V_{,\phi} = 0,
\end{equation}
while the neutrino energy density evolves according to the standard radiation continuity equation,
\begin{equation}
\dot{\rho}_\nu + 4H\rho_\nu = 0.
\end{equation}
In this regime, the effective potential  \eqref{2.13} reduces to
\begin{equation}
V_{\mathrm{eff}} = V,
\end{equation}
showing that relativistic neutrinos do not influence the scalar-field evolution.

At late times, neutrinos become non-relativistic, satisfying \( m_\nu \gg T_\nu \). In this regime, their pressure becomes negligible, \( P_\nu \approx 0 \), and the trace of the energy-momentum tensor is dominated by the energy density, \( \rho_\nu - 3P_\nu \approx \rho_\nu \). Consequently, the scalar--neutrino coupling becomes dynamically important. The Klein--Gordon equation~\eqref{2.13} and the continuity equation~\eqref{2.11} then take the form
\begin{equation}\label{2.19}
\ddot{\phi} + 3H\dot{\phi} + V_{,\phi}
=
- \frac{m_{\nu,\phi}}{m_\nu}\rho_\nu,
\end{equation}
and
\begin{equation}\label{2.20}
\dot{\rho}_\nu + 3H\rho_\nu
=
\frac{m_{\nu,\phi}}{m_\nu}\dot{\phi}\rho_\nu,
\end{equation}
respectively. Eq.\eqref{2.20} shows that the neutrino energy density is no longer independently conserved, but exchanges energy with the scalar field due to the time dependence of $m_\nu(\phi)$.
It is convenient to introduce a rescaled neutrino energy density $\hat{\rho}_\nu$ through
\begin{equation}
    \rho_\nu = m_\nu \hat{\rho}_\nu.
\end{equation}
In terms of $\hat{\rho}_\nu$, Eqs.~\eqref{2.19} and \eqref{2.20} become 
\begin{equation}\label{2.22}
\begin{aligned}
\ddot{\phi} + 3H\dot{\phi} + V_{,\phi}
+ m_{\nu,\phi}\hat{\rho}_\nu &= 0, \\
\dot{\hat{\rho}}_\nu + 3H\hat{\rho}_\nu &= 0 .
\end{aligned}
\end{equation}
In the non-relativistic limit one may write $\rho_\nu = m_\nu n_\nu$, where $n_\nu$ is the neutrino number density, which can be identified with $\hat{\rho}_\nu$. On solving the second equation in \eqref{2.22} gives
\begin{equation}
    \hat{\rho}_\nu = \hat{\rho}_{\nu 0} a^{-3},
\end{equation}
where the subscript ''0'' denoted the present epoch and $a_0=1$.
Accordingly, in the non-relativistic regime the effective potential \eqref{2.22} is written as,
\begin{equation}
    V_{\mathrm{eff},\phi}
=
V_{,\phi}
+
m_{\nu,\phi} n_\nu.
\end{equation}
This explicitly shows that once neutrinos become non relativistic, they contribute an additional $\phi$-dependent term to the effective potential, which may induce a minimum in $V_{\mathrm{eff}}$ and significantly modify the late-time cosmological dynamics.

In the non-relativistic neutrino regime, the background dynamics of a spatially flat Friedmann-Lemaître-Robertson-Walker (FLRW) Universe, comprising dark energy, non-relativistic neutrinos, and pressureless matter (baryons), are governed by
\begin{align}
&H^2 = \frac{8\pi G}{3}\left(\rho_\phi + \rho_\nu + \rho_m \right), 
\label{fried1} \\
&\dot H + H^2 = -\frac{4\pi G}{3} 
\left(\rho_\phi + \rho_\nu + \rho_m + 3 p_\phi \right),
\label{fried2}
\end{align}
where $\rho_\phi$, $\rho_\nu$, and $\rho_m$ denote the energy 
densities of the scalar field (dark energy), neutrinos, and 
pressureless matter (including baryons and cold dark matter), 
respectively. The energy density and pressure for the dark energy that is described by the scalar field $\phi$ are given as 
\begin{align}
\rho_\phi= \frac{1}{2}\dot{\phi}^2 + V(\phi), \qquad
p_\phi = \frac{1}{2}\dot{\phi}^2 - V(\phi).
\end{align}
where $V(\phi)$ denotes the scalar field potential. The continuity equation for neutrino species can be written in the form
\begin{equation}\label{neutrino}
    \dot\rho_\nu+3H\rho_\nu= \delta(a)H\rho_\nu,
\end{equation}
where, $\delta(a)$ represents the coupling function. In comparison with Eq.(\ref{2.11}), we define the coupling function as
\begin{equation}
    \delta(a)=\frac{d \ln{m_\nu}}{d \ln{a}}
\end{equation}
Also, the energy density of pressurless matter and dark energy should obey 
\begin{align}
\dot \rho_m+3H\rho_m&=0\label{bary}\\
\dot\rho_\phi+3H\rho_\phi(1+\omega_\phi)&=-\delta(a)H\rho_\nu\label{phi}.\end{align}
To evaluate the energy densities of different components, we work with the different variable defined, $\alpha=\ln{a}=-\ln{(1+z)}$. Using the different variable $\alpha$, we obtain the solution for the continuity given in Eq.(\ref{neutrino}) in terms of redshift $z$ as
\begin{equation}
\rho_\nu(z)=\rho_\nu^0 (1+z)^3 
\exp\left[-\int_{0}^{z}\frac{\delta(z')}{1+z'}\,dz'\right]
\end{equation}
To evaluate the energy density of the dark energy component, we write the continuity equation in the variable $\alpha$ as
\begin{equation}\label{rhophiused}
\rho'_\phi(\alpha)+3\rho_\phi(1+\omega_\phi(\alpha))=-\delta(\alpha)\rho_\nu,
\end{equation}
where $\rho'_\phi\equiv \frac{d\rho_\phi}{d\alpha}$.
The above equation can be written in the form of an effective fluid equation 
\begin{equation}
 \rho'_\phi(\alpha)+3\rho_\phi(1+\omega^{eff}_\phi(\alpha))=0  
\end{equation}
and now comparing it with the ideal dark energy continuity equation yields the effective equation of state 
\begin{equation}
    \omega^{eff}_\phi= \omega_\phi + \frac{\delta(\alpha) \rho_\nu}{3\rho_\phi}
\end{equation}
Now, to evaluate the solution energy density of dark energy, we solve the  differential equation Eq.(\ref{rhophiused})
will provide a general solution 
\begin{equation}
\rho_\phi(\alpha)= \frac{1}{e^{\int 3(1+\omega_\phi(\alpha))d\alpha}}\left[-\int e^{\int 3(1+\omega_\phi(\alpha))d\alpha}(\delta(\alpha) \rho_\nu)d\alpha +C\right],
\end{equation}
In terms of redshift,
\begin{equation}\label{fz}
\rho_\phi(z)=
\frac{1}{\exp\!\left[-\int \frac{3(1+\omega_\phi(z))}{1+z}\,dz\right]}
\left[
\int
\exp\!\left(-\int \frac{3(1+\omega_\phi(z))}{1+z}\,dz\right)
\delta(z)\rho_\nu\,\frac{dz}{1+z}
+ C
\right].
\end{equation}
where $C$ is the integration constant evaluated when $\rho_\phi(\alpha = 0) = \rho^0_\phi$, with $\rho_\phi^0$
being the dark
energy density today.
Using the Friedmann equation and from different energy density equations, we write the normalized Hubble equation as
\begin{equation}
    \frac{H^2(z)}{H_0^2}
=
\Omega_m (1+z)^3
+
\Omega_\nu (1+z)^3
\exp\!\left[-
\int_0^z
\frac{\delta(z')}{1+z'}\, dz'
\right]
+
\Omega_\phi f(z)
\end{equation}
where $f(z)$ encodes the evolution of the dark energy component and is determined by solving Eq.~(\ref{fz}) numerically. The evolution equation of the scalar field is obtained as
\begin{equation}
    \left(\frac{d\phi}{dz}\right)^2
=
\frac{1}{4\pi G (1+z)^2}
\left[
(1+z)\frac{d\ln H}{dz}
-
\frac{3}{2}\left(\Omega_\nu + \Omega_m\right)
\right].
\end{equation}
The dynamical evolution of the system is governed by the equation-of-state parameter $\omega$ and the coupling function $\delta$. Therefore, in the following subsections, we perform a case-by-case analysis of the evolution equations for different choices of $\omega$ and $\delta$.

\subsection{Case I: Constant $\omega_\phi$ and $\delta(z)$}
For the special case, where $\omega_\phi=\omega=\text{constant}$ and $\delta=\text{constant}$, the energy density of neutrino and dark energy are calculated as
\begin{align}
\rho_\nu(z)&=\rho_\nu^0 (1+z)^{3-\delta},\\ \vspace{0.8cm}
\rho_\phi(z)&= \rho^0_\phi (1+z)^{3(1+\omega)} 
+ \frac{\delta\rho^0_\nu}{3\omega+\delta}
\left[(1+z)^{3(1+\omega)}-(1+z)^{3-\delta}\right].
\end{align}
For constant $\delta$ and constant equation-of-state parameter $\omega$, the normalized Hubble parameter in terms of redshift is given by
\begin{equation}
\frac{H^2(z)}{H_0^2}
= \Omega_m (1+z)^3
+ \Omega_\nu (1+z)^{3-\delta}
+ \Omega_\phi f(z),
\end{equation}
where the dark energy evolution function $f(z)$ takes the form
\begin{equation}
f(z)
= (1+z)^{3(1+\omega)}
+ \frac{\delta\,\Omega_\nu}{\Omega_\phi(3\omega+\delta)}
\left[(1+z)^{3(1+\omega)} - (1+z)^{3-\delta}\right].
\end{equation}

\subsection{Case II: Constant $\omega_\phi$ and variable $\delta(z)$}
For the case of a variable coupling function $\delta(z)$, we consider two representative forms of mass-varying neutrino models. The first corresponds to a purely exponential dependence on the scalar field, given by $m_\nu(\phi) = m_{\nu0} e^{\beta \phi}$ with the exponential potential $
V(\phi)=V_0e^{-\lambda\phi}$. For a constant equation of state $\omega_\phi=\omega_0$, the energy density of the neutrino for an exponential mass term is given as 
\begin{equation}
\rho_\nu= \rho^0_\nu e^{-3\alpha} e^{\beta(\phi(\alpha)-\phi_0)}
\end{equation}

For the exponential mass-varying neutrino model, we consider the neutrino mass to depend on the scalar field as

\begin{equation}
m_\nu(\phi)=m_{\nu0}e^{\beta\phi},
\end{equation}

In general, the coupled continuity equations and the energy density Eq.(\ref{fz}) do not admit a closed-form analytical solution for the scalar-field energy density. Consequently, a direct evaluation of $\rho_\phi$ requires solving the full dynamical system numerically.

To obtain an analytical description suitable for observational analyses, we adopt the adiabatic approximation commonly employed in mass-varying neutrino cosmologies\cite{Brookfield2006PRD,Brookfield2006}. In this regime, the scalar field rapidly relaxes to the instantaneous minimum of the effective potential 
\begin{equation}
V_{\rm eff}(\phi)
=V(\phi)
+
\rho_{\nu0}a^{-3}
e^{\beta(\phi-\phi_0)},
\end{equation}
The validity of this approximation may be understood by expanding the scalar field about the minimum,

\begin{equation}
\phi=\phi_{\rm min}+\delta\phi,
\end{equation}
which reduces the Klein-Gordon equation to
\begin{equation}
\delta\ddot{\phi}
+
3H\delta\dot{\phi}
+
m_{\rm eff}^2\delta\phi
=
0,\quad 
\text{where},
\quad
m_{\rm eff}^2
\equiv
V_{{\rm eff},\phi\phi}
(\phi_{\rm min})
\end{equation}
``$m_{\rm eff}$" is the effective mass of the scalar field. For the condition,
\begin{equation}
m_{\rm eff}^2 \gg H^2,
\end{equation}
the oscillation timescale of the field is much shorter than the Hubble timescale, and Hubble friction efficiently damps the oscillations around the minimum. The scalar field therefore adiabatically tracks the instantaneous minimum of the effective potential, which satisfies
\begin{equation}
\frac{\partial V_{\rm eff}}{\partial \phi}=0.
\end{equation}
this gives
\begin{equation}
\phi(z)
= \phi_*-\frac{3}{\beta+\lambda}\ln(1+z),
\end{equation}
where
\begin{equation}
\phi_*
=\frac{1}{\beta+\lambda}
\left[
\ln\left(
\frac{\lambda V_0}{\beta\rho_{\nu}^0}
\right)
+\beta\phi_0
\right].
\end{equation}
Consequently, the effective interaction parameter is given by
\begin{equation}\label{det}
\delta_{\rm eff}=\frac{3\beta}{\beta+\lambda}.
\end{equation}
The energy densities of neutrinos and dark energy are then obtained as
\begin{align}
\rho_\nu(z)
&=
\rho_\nu^0 (1+z)^{3-\delta_{\rm eff}},\\
\rho_\phi(z)
&=
\rho_\phi^0 (1+z)^{3(1+\omega_0)}
+
\frac{\delta_{\rm eff}\rho_\nu^0}{3\omega_0+\delta_{\rm eff}}
\left[
(1+z)^{3(1+\omega_0)}
-(1+z)^{3-\delta_{\rm eff}}
\right].
\end{align}
We therefore obtained the same functional structure as the interacting dark-energy model discussed in Case I. The crucial difference is that the interaction parameter is no longer introduced phenomenologically but is instead determined through Eq.(\ref{det}) which depends on the parameters $\beta$ and $\lambda$ of potential mass of MaVaN.
For an exponential mass-varying neutrino model and a constant $\omega_0$, the normalized Hubble parameter in terms of redshift is given by

\begin{equation}
\frac{H^2(z)}{H_0^2}
=
\Omega_m (1+z)^3
+
\Omega_\nu (1+z)^{3-\delta_{\rm eff}}
+
\Omega_\phi f(z),
\end{equation}
where the dark-energy evolution function takes the form
\begin{equation}
f(z)
=(1+z)^{3(1+\omega_0)}
+
\frac{\delta_{\rm eff}\Omega_\nu}
{\Omega_\phi(3\omega_0+\delta_{\rm eff})}
\left[
(1+z)^{3(1+\omega_0)}
-(1+z)^{3-\delta_{\rm eff}}
\right].
\end{equation}
\subsection{Case III: Variable $\omega_\phi$ and constant $\delta(z)$}
For the variable equation of state, we choose the CPL parametrization 
\begin{equation}
\omega_\phi = w_0 + w_1(1-a)
\end{equation}
and constant coupling $\delta$, the scalar field energy density becomes
\begin{equation}
\rho_\phi(z)
=
(1+z)^{3(1+w_0+w_1)}
\exp\left[\frac{3w_1}{1+z}\right]
\left[
-\delta \rho_\nu^0 
\int (1+z)^{-1-3w_0-3w_1}
\exp\left(-\frac{3w_1}{1+z'}\right) dz'
+ C
\right].
\end{equation}Imposing the boundary condition $\rho_\phi(0)=\rho_{\phi0}$.
The dark-energy evolution function becomes
\begin{equation}
f(z)
=
(1+z)^{3(1+w_0+w_1)}
\exp\left(\frac{3w_1}{1+z}\right)
\left[
1
-
\frac{\delta\Omega_\nu}{\Omega_\phi}
\int_0^z
(1+z')^{-1-3w_0-3w_1}
\exp\left(-\frac{3w_1}{1+z'}\right)
dz'
\right].
\end{equation}

\begin{equation}
\begin{aligned}
\rho_\phi(z)
=&\,
(1+z)^3
e^{\frac{3w_1}{1+z}}
\Bigg[
(1+z)^{3(w_0+w_1)}
\left(
\rho_{\phi0}e^{-3w_1}
-\delta\rho_{\nu0}
E_{1-3(w_0+w_1)}
\!\left(3w_1\right)
\right)
\\
&
\qquad\qquad
+\delta\rho_{\nu0}
E_{1-3(w_0+w_1)}
\!\left(
\frac{3w_1}{1+z}
\right)
\Bigg].
\end{aligned}
\end{equation}
where $( \rho_{\phi0} )$ and $( \rho_{\nu0} )$ denote the present-day dark-energy and neutrino energy densities, respectively, and
\begin{equation}
E_n(x)=\int_{1}^{\infty}
\frac{e^{-xt}}{t^n}\,dt,
\qquad \Re(x)>0,
\end{equation}
is the generalized exponential integral function. It is related to the upper incomplete gamma function through
\begin{equation}
E_n(x)=x^{\,n-1}\Gamma(1-n,x),
\end{equation}
where
\begin{equation}
\Gamma(s,x)
=
\int_x^\infty t^{s-1}e^{-t}\,dt
\end{equation}
denotes the upper incomplete gamma function.

The Normalized Hubble equation 
\begin{equation}
\frac{H^2(z)}{H_0^2}
= \Omega_m (1+z)^3
+ \Omega_\nu (1+z)^{3-\delta}
+ \Omega_\phi f(z),
\end{equation}

where 
\begin{equation}
\begin{aligned}
f(z)
=&\,
e^{-\frac{3w_1 z}{1+z}}
(1+z)^3
\Bigg[
(1+z)^{3(w_0+w_1)}
\Bigg(
1
-
e^{3w_1}
\frac{\delta\,\Omega_{\nu0}}
{\Omega_{\phi0}}
E_{1-3(w_0+w_1)}
\!\left(3w_1\right)
\Bigg)
\\
&
\qquad\qquad
+
e^{3w_1}
\frac{\delta\,\Omega_{\nu0}}
{\Omega_{\phi0}}
E_{1-3(w_0+w_1)}
\!\left(
\frac{3w_1}{1+z}
\right)
\Bigg].
\end{aligned}
\end{equation}

\section{Dataset and Methodology}\label{DATASET}
To constrain the parameters of the model under investigation, we employ the \texttt{emcee} package \cite{foreman2013emcee}, an affine-invariant ensemble sampler implementation of the Markov Chain Monte Carlo (MCMC) algorithm, to explore the posterior parameter space of the cosmological model. We perform four independent MCMC analyses, combining the DESI DR2 BAO measurements and the compressed CMB likelihood with each of the following Type Ia supernova compilations in turn: Pantheon$^{+}$, DES-Dovekie, and Union3, in addition to a baseline analysis using only DESI DR2 BAO and the compressed CMB likelihood without any supernova sample. The convergence of the MCMC chains is assessed using the integrated autocorrelation time $\tau$, estimated directly from the chains. We require the total chain length to exceed $50\,\tau_{\max}$, where $\tau_{\max}$ is the largest autocorrelation time among all sampled parameters, following the standard convergence criterion recommended for \texttt{emcee}~\cite{foreman2013emcee}. The first $3\,\tau_{\max}$ steps of each chain are discarded as burn-in, and the remaining samples are thinned by a factor of $\tau_{\max}/2$ to reduce residual autocorrelation before further analysis. Once the final chains are obtained, we use the \texttt{GetDist} package~\cite{lewis2025getdist} for post-processing and visualization. To constrain the parameters of the MaVaN cosmological model, we compare the model against Baryon Acoustic Oscillation (BAO) measurements from the Dark Energy Spectroscopic Instrument Data Release 2 (DESI DR2), Type Ia supernova observations, and the compressed CMB likelihood, which are detailed below.
\begin{itemize}
     \item \textbf{Baryon Acoustic Oscillation :} First, we use recent Baryon Acoustic Oscillation (BAO) measurements from the DESI Data Release 2 (DR2) \cite{karim2025desi}. These measurements are extracted using various tracers such as the Bright Galaxy Sample (BGS), Luminous Red Galaxies (LRG1–3), Emission Line Galaxies (ELG1–2), Quasars (QSO), and Lyman-$\alpha$ forests. To use these measurements, we compute the Hubble distance $D_H(z) = \frac{c}{H(z)}$, the comoving angular diameter distance $D_M(z) = c \int_0^z \frac{dz'}{H(z')}$, and the volume-averaged distance $D_V(z) = \left[ z, D_M^2(z), D_H(z) \right]^{1/3}$. It is necessary to derive the following ratios: $D_M/r_d$, $D_H/r_d$, $D_V/r_d$, and $D_M/D_H$ to constrain the parameters of each model, where $r_d$ is the sound horizon. In flat $\Lambda$CDM, $r_d = 147.09 \pm 0.20$ Mpc \cite{aghanim2020planck}.
     \item \textbf{Type Ia Supernova :} Then, we use three different Type Ia supernova (SNe Ia) measurements. The first is the Pantheon$^{+}$ sample \cite{brout2022pantheon}, which includes 1,701 light curves from 1,550 SNe Ia observations spanning the redshift range $0.001 \leq z \leq 2.26$. In our analysis, we exclude light curves at $z < 0.01$, as such low-redshift data are affected by significant systematic uncertainties due to peculiar velocities. Second, we use 1,829 photometric light curves spanning the redshift range $0.10 \leq z \leq 1.13$, collected over five years by the Dark Energy Survey Supernova Program (DES-SN5Y)~\cite{abbott2024dark}. This sample includes 1,635 DES discovered Type~Ia supernovae and 194 externally sourced low-$z$ supernovae from the CfA and CSP samples. Finally, we use the Union3 compilation, which contains 2,087 cosmologically useful Type~Ia supernovae from 24 datasets covering the redshift range $0.05 \leq z \leq 2.26$~\cite{rubin2025union}.
     \item \textbf{Compressed CMB likelihood :} Finally, we use the compressed CMB likelihood, characterized by the parameter vector $\mathbf{v} = {\theta_s^{-1}(z_\ast), \omega_b, \omega_c}$, which is modeled as a 3×3 Gaussian likelihood~\cite{aubourg2015cosmological} and implemented as the \texttt{PLK18} likelihood within \texttt{SimpleMC}. We use this compressed likelihood because the Anton Schmidt model mainly affects the late time expansion history of the Universe. The full CMB power spectrum shows small non-geometric features, such as an enhanced lensing amplitude and a low-$\ell$ power deficit, which may be caused by residual systematics and can bias dark energy constraints. For instance, Planck data alone show a $\gtrsim 2\sigma$ preference for phantom-like dark energy~\cite{escamilla2024state}, largely due to the lack of large-scale power. To avoid such biases, we adopt the compressed CMB likelihood in our analysis.
\end{itemize}

\section{Results and Discussion}

This section presents the constraints obtained for the three neutrino--dark-energy coupling scenarios introduced in Sec.~\ref{b}: a constant coupling with a constant dark-energy equation of state (Case~I), an exponential mass-varying neutrino coupled to a quintessence field (Case~II), and a constant coupling combined with a CPL dark-energy background (Case~III). Each case is fit to DESI~DR2 BAO and a compressed CMB likelihood, combined in turn with no supernova sample, Pantheon$^{+}$, DES-Dovekie, and Union3, so the same four dataset combinations can be compared across all three cases. This lets us separate what the data prefer from what follows only from the way the coupling is parametrized.

\subsection{Case I: Constant $\omega_\phi$ and $\delta$}\label{sec:results_case1}

Table~\ref{tab:case1_results} lists the posterior constraints on the five Case~I parameters, $\{H_0,\Omega_{m0},\Omega_{\nu0},\omega,\delta\}$, together with the derived total neutrino mass, $\sum m_\nu = 93.14\,\Omega_{\nu0}h^2\,\mathrm{eV}$, for the four dataset combinations described above.

\begin{table}[H]
\centering
\begin{tabular}{lcccc}
\hline\hline
Parameter & CMB+DESI DR2 & +Pantheon$^{+}$ & +DES-Dovekie & +Union3 \\
\hline
$H_0\ [\mathrm{km\,s^{-1}\,Mpc^{-1}}]$ & $71.07^{+0.73}_{-0.73}$ & $69.18^{+0.49}_{-0.50}$ & $68.62^{+0.47}_{-0.47}$ & $67.74^{+0.59}_{-0.60}$ \\
$\Omega_{m0}$                          & $0.2786^{+0.0067}_{-0.0066}$ & $0.2957^{+0.0051}_{-0.0048}$ & $0.3011^{+0.0049}_{-0.0048}$ & $0.3087^{+0.0059}_{-0.0058}$ \\
$\Omega_{\nu0}$                        & $0.00023^{+0.00051}_{-0.00019}$ & $0.00032^{+0.00065}_{-0.00025}$ & $0.0004^{+0.0007}_{-0.0003}$ & --- \\
$\omega$                               & $-1.105^{+0.033}_{-0.034}$ & $-1.016^{+0.022}_{-0.022}$ & $-0.9894^{+0.0212}_{-0.0216}$ & $-0.958^{+0.027}_{-0.027}$ \\
$\delta$                               & $-0.201^{+0.147}_{-0.280}$ & $-0.221^{+0.163}_{-0.292}$ & $-0.2171^{+0.1591}_{-0.2988}$ & $-0.249^{+0.181}_{-0.326}$ \\
$\sum m_\nu\ [\mathrm{eV}]$ (95\% CL)  & $<0.06$ & $<0.07$ & $<0.08$ & $<0.18$ \\
\hline\hline
\end{tabular}\caption{Marginalized posterior constraints (median with $68\%$ credible intervals) on the Case~I ($\omega_\phi=\text{const}$, $\delta=\text{const}$) parameters, for the four dataset combinations analyzed in this work. The total neutrino mass $\sum m_\nu$, a derived parameter, is quoted as a $95\%$ CL upper limit.}
\label{tab:case1_results}
\end{table}

Adding a supernova sample to CMB+DESI~DR2 lowers $H_0$ steadily, from $71.07^{+0.73}_{-0.73}$ to $67.74^{+0.59}_{-0.60}\,\mathrm{km\,s^{-1}\,Mpc^{-1}}$ once Union3 is included, while $\Omega_{m0}$ rises from $0.2786^{+0.0067}_{-0.0066}$ to $0.3087^{+0.0059}_{-0.0058}$ over the same sequence -- the usual $H_0$--$\Omega_{m0}$ degeneracy of a flat, distance-based fit, since a lower $H_0$ needs a larger $\Omega_{m0}$ to keep the same distances. The CMB+DESI~DR2-only fit sits at one end of this trend, with the highest $H_0$, the lowest $\Omega_{m0}$, and, as discussed next, the equation of state farthest from $-1$.

$\omega$ shows the largest change of any parameter in the table. With CMB+DESI~DR2 alone, $\omega=-1.105^{+0.033}_{-0.034}$ lies in phantom territory, more than $3\sigma$ from $-1$; adding a supernova sample moves it back: $-1.016^{+0.022}_{-0.022}$ with Pantheon$^{+}$, $-0.9894^{+0.0212}_{-0.0216}$ with DES-Dovekie, and $-0.958^{+0.027}_{-0.027}$ with Union3, about $1.5\sigma$ from $-1$. A similar pattern -- BAO+CMB preferring $\omega<-1$, with supernovae pulling the fit back across $-1$ -- has been reported in other DESI-DR2-based dynamical dark-energy analyses~\cite{karim2025desi,escamilla2024state}. Here it can be traced to the coupling term in Eq.~(\ref{rhophiused}), which turns the bare $\omega$ into an effective $\omega_\phi^{\rm eff}$, together with the fact that BAO and supernovae probe different redshift ranges.

The coupling $\delta$ changes little across the four combinations, staying within $[-0.249,-0.201]$ with heavily overlapping intervals. Since $\rho_\nu(z)\propto(1+z)^{3-\delta}$, a negative $\delta$ dilutes the neutrino density faster than a standard species would -- consistent with energy flowing from neutrinos into the scalar field at late times, the mechanism linked in Sec.~\ref{b} to the onset of acceleration. The posterior leans toward more negative values in every case, but $\delta=0$ stays inside the $68\%$ interval throughout, so a coupling is favored without being required. The derived neutrino mass upper limit moves together with $\Omega_{m0}$ and $H_0$, tightening to $\sum m_\nu<0.06\,\mathrm{eV}$ (CMB+DESI~DR2) and loosening to $\sum m_\nu<0.18\,\mathrm{eV}$ (Union3); in every case a massless or nearly massless neutrino remains fully consistent with the data.

\begin{figure}[H]
\centering
\includegraphics[width=0.85\textwidth]{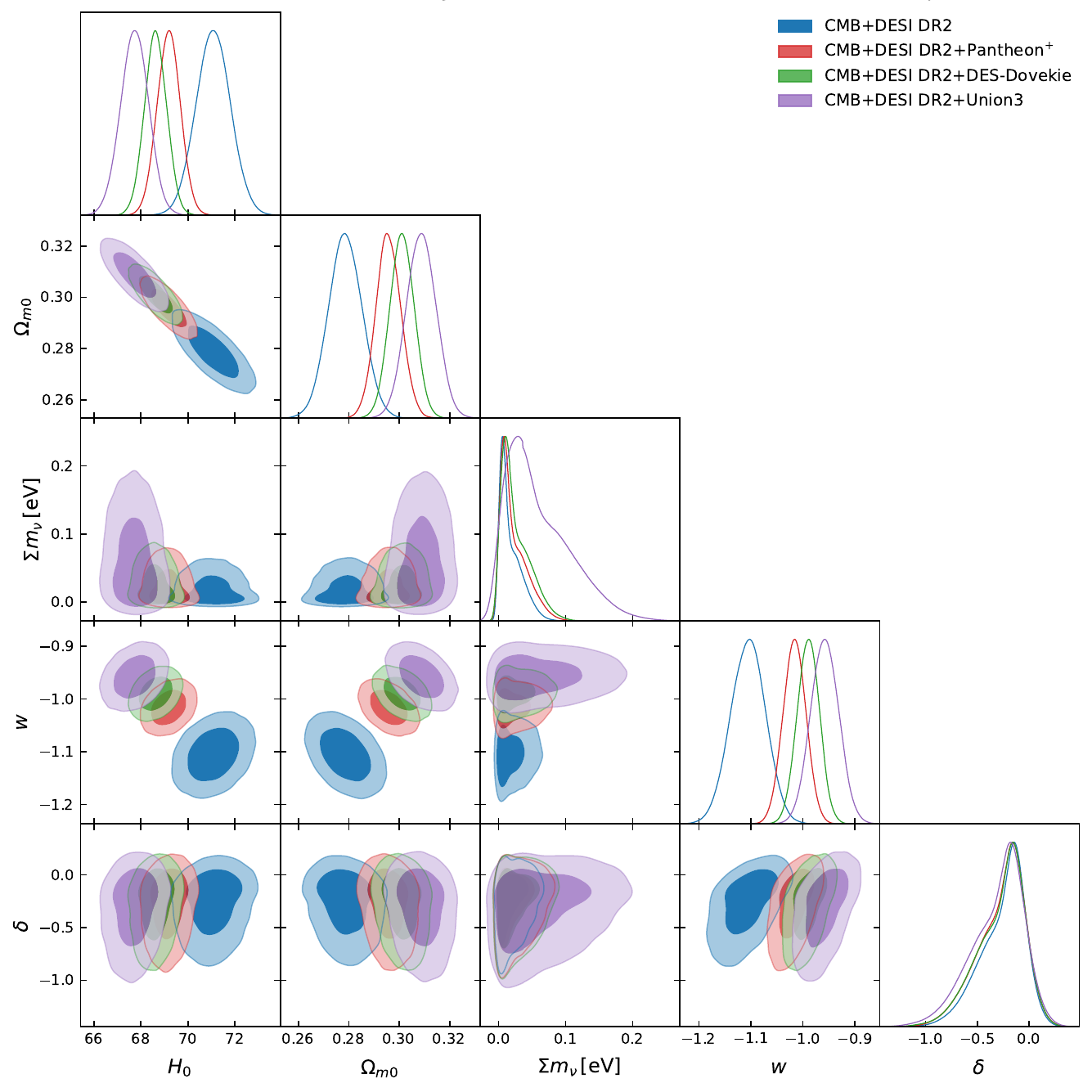}
\caption{Marginalized one-dimensional posterior distributions (diagonal panels) and two-dimensional $68\%$ and $95\%$ credible regions (off-diagonal panels) for the Case~I parameters $\{H_0,\,\Omega_{m0},\,\sum m_\nu,\,\omega,\,\delta\}$, obtained from the joint analysis of DESI DR2 BAO, DES-SN5YR, and the compressed CMB early-Universe prior.}
\label{fig:corner_case1}
\end{figure}

Figure~\ref{fig:corner_case1} shows a strong $H_0$--$\Omega_{m0}$ anticorrelation, a weaker $\Omega_{m0}$--$\omega$ anticorrelation, and a moderate positive $\omega$--$\delta$ correlation. The last of these follows from $\omega$ and $\delta$ appearing together in the denominator $(3\omega+\delta)$ of the dark-energy evolution function derived in Sec.~\ref{b}, so the two are not fully separable with these data. The $\sum m_\nu$ posterior is non-Gaussian in every fit, rising near zero and trailing off in a long tail -- the shape expected when an analysis is reporting an upper limit rather than a detection.

\subsection{Case II: Exponential Mass-Varying Neutrino}\label{sec:results_case2}

In Case~II the neutrino mass tracks the scalar field exponentially, $m_\nu(\phi)=m_{\nu0}e^{\beta\phi}$, with potential $V(\phi)=V_0e^{-\lambda\phi}$. Under the adiabatic minimum-tracking approximation of Sec.~\ref{b}, this reduces to an effective, redshift-independent coupling
\begin{equation}
\delta_{\rm eff}=\frac{3\beta}{\beta+\lambda},
\end{equation}
so $\beta$ and $\lambda$, rather than $\delta_{\rm eff}$ itself, are the quantities sampled directly; $\delta_{\rm eff}$ and $\sum m_\nu$ are derived and inherit whatever priors were placed on $\beta$ and $\lambda$. Table~\ref{tab:case2_results} lists the resulting constraints for the same four dataset combinations as Case~I.

\begin{table}[H]
\centering
\begin{tabular}{lcccc}
\hline\hline
Parameter & CMB+DESI DR2 & +Pantheon$^{+}$ & +DES-Dovekie & +Union3 \\
\hline
$H_0\ [\mathrm{km\,s^{-1}\,Mpc^{-1}}]$ & $68.29^{+0.73}_{-0.73}$ & $68.14^{+0.49}_{-0.50}$ & $69.64^{+0.21}_{-0.21}$ & $67.81^{+0.58}_{-0.59}$ \\
$\Omega_{m0}$                          & $0.3036^{+0.0071}_{-0.0070}$ & $0.3051^{+0.0050}_{-0.0049}$ & $0.2893^{+0.0022}_{-0.0022}$ & $0.3083^{+0.0059}_{-0.0058}$ \\
$\beta$                                & $0.050^{+0.107}_{-0.148}$ & $-0.153^{+0.111}_{-0.186}$ & $0.064^{+0.126}_{-0.417}$ & $0.041^{+0.114}_{-0.148}$ \\
$\lambda$                              & $2.990^{+0.685}_{-0.682}$ & $2.546^{+0.312}_{-0.357}$ & $2.838^{+0.772}_{-0.601}$ & $2.996^{+0.679}_{-0.677}$ \\
$\delta_{\rm eff}$            & $0.050^{+0.101}_{-0.151}$ & $-0.196^{+0.144}_{-0.273}$ & $0.538^{+0.216}_{-0.324}$ & $0.040^{+0.108}_{-0.151}$ \\
$\omega$                               & $-0.990^{+0.033}_{-0.034}$ & $-0.966^{+0.022}_{-0.023}$ & $-0.960^{+0.015}_{-0.015}$ & $-0.964^{+0.027}_{-0.027}$ \\
$\sum m_\nu\ [\mathrm{eV}]$ (95\% CL)  & $<0.32$ & $<0.14$ & $<0.19$ & $<0.31$ \\
$M_B$                                  & --- & $-19.403^{+0.012}_{-0.012}$ & --- & --- \\
\hline\hline
\end{tabular}
\caption{Marginalized posterior constraints (median with $68\%$ credible intervals) on the Case~II parameters, for the four dataset combinations analyzed in this work. $\delta_{\rm eff}$ is a derived quantity; $\sum m_\nu$ is also derived and is quoted as a $95\%$ CL upper limit; $M_B$ is an additional nuisance parameter fit only in the Pantheon$^{+}$ analysis (see text).}
\label{tab:case2_results}
\end{table}

$H_0$ and $\Omega_{m0}$ again move in opposite directions, but the ordering across datasets differs from Case~I: DES-Dovekie now gives the highest $H_0$, $69.64^{+0.21}_{-0.21}\,\mathrm{km\,s^{-1}\,Mpc^{-1}}$, above the CMB+DESI~DR2-only value of $68.29^{+0.73}_{-0.73}$, while Pantheon$^{+}$ and Union3 pull it down to $68.14^{+0.49}_{-0.50}$ and $67.81^{+0.58}_{-0.59}$; $\Omega_{m0}$ tracks this inversely. The equation of state also behaves differently: CMB+DESI~DR2 alone now gives $\omega=-0.990^{+0.033}_{-0.034}$, consistent with $-1$, while the other three combinations stay close by, between $-0.966$ and $-0.960$. The phantom-like value found for BAO+CMB alone in Case~I is therefore not a fixed feature of the data -- it depends on how the coupling is parametrized, and here it is absorbed once $\beta$ and $\lambda$ are free to adjust.

$\delta_{\rm eff}$ is less stable than $\delta$ was in Case~I: consistent with zero for CMB+DESI~DR2 ($0.050^{+0.101}_{-0.151}$) and Union3 ($0.040^{+0.108}_{-0.151}$), positive for DES-Dovekie ($0.538^{+0.216}_{-0.324}$), and negative for Pantheon$^{+}$ ($-0.196^{+0.144}_{-0.273}$). Part of this spread has a simple explanation: the four runs did not share the same priors on $\beta$ and $\lambda$ -- CMB+DESI~DR2 and Union3 used $\beta\in[-0.2,0.2]$, $\lambda\in[2,4]$; DES-Dovekie widened this to $\beta\in[-1,1]$; Pantheon$^{+}$ restricted $\beta\in[-1,0]$ and $\lambda\in[2,3]$. Since $\delta_{\rm eff}$ depends nonlinearly on $\beta$ and $\lambda$, a difference in prior volume alone can shift or flip its sign, so the negative $\delta_{\rm eff}$ from Pantheon$^{+}$ should not be read as an independent confirmation of the coupling found in Case~I: that run was never allowed to explore $\beta>0$.\footnote{The Pantheon$^{+}$ run also fits the absolute magnitude $M_B=-19.403^{+0.012}_{-0.012}$ directly rather than marginalizing over distance moduli as in the other three runs -- a separate methodological difference worth resolving for a fully consistent comparison.} A uniform, sufficiently wide prior on $\beta$ and $\lambda$ across all four runs (e.g., $\beta\in[-1,1]$, $\lambda\in[2,4]$) would put them on equal footing. The derived $\sum m_\nu$ $95\%$ CL upper limits range from $<0.14\,\mathrm{eV}$ (Pantheon$^{+}$) to $<0.32\,\mathrm{eV}$ (CMB+DESI~DR2, matched by Union3), noticeably looser than in Case~I; since $\sum m_\nu$ is itself reconstructed nonlinearly from $\beta$ and $\lambda$, part of this loosening may also reflect the differing prior volumes rather than the data, and would benefit from the same unification.

\begin{figure}[H]
\centering
\includegraphics[width=0.85\textwidth]{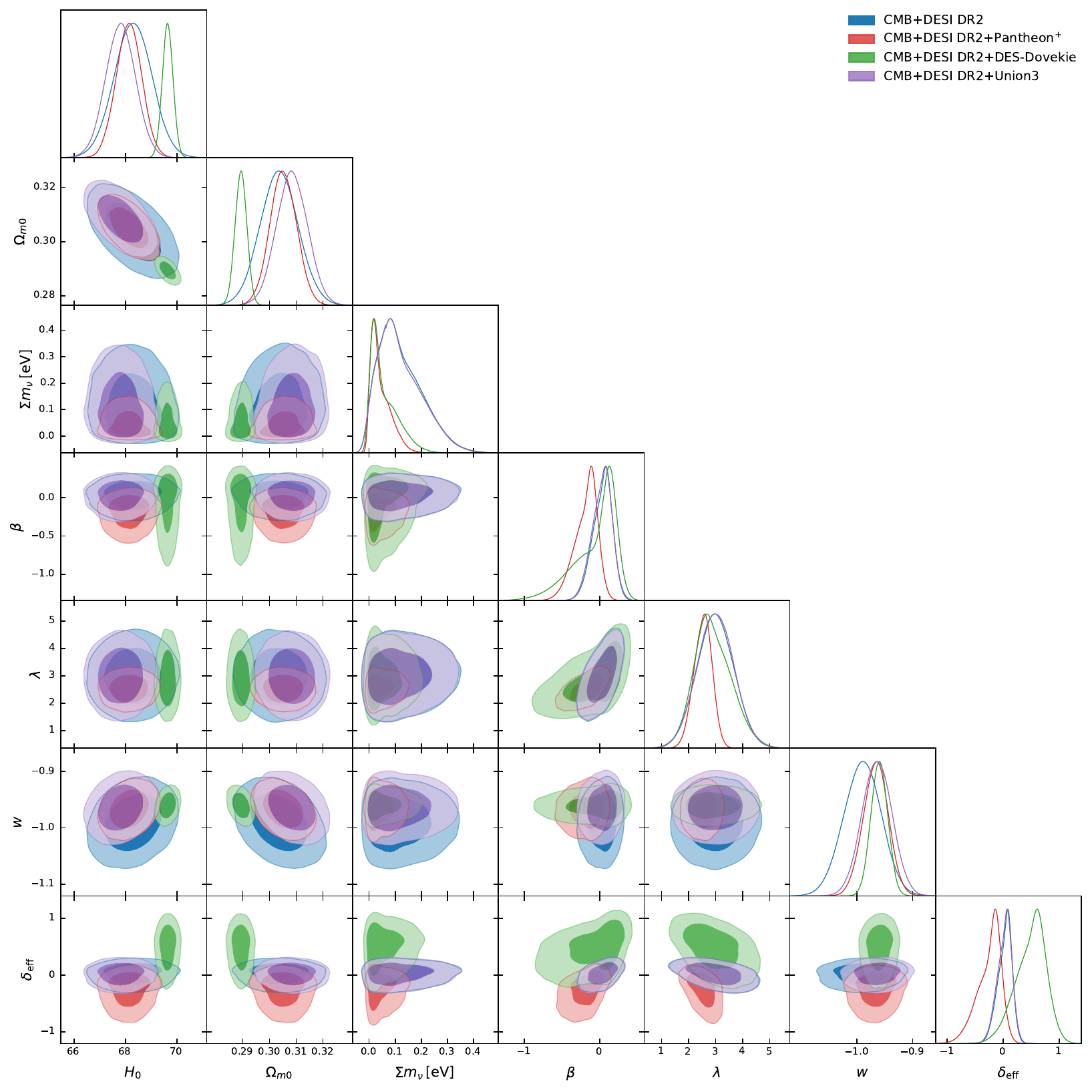}
\caption{Marginalized one-dimensional posterior distributions (diagonal panels) and two-dimensional $68\%$ and $95\%$ credible regions (off-diagonal panels) for the Case~II parameters $\{H_0,\,\Omega_{m0},\,\beta,\,\lambda,\,\omega\}$, for the four dataset combinations of Table~\ref{tab:case2_results}. Note the differing prior ranges adopted for $\beta$ and $\lambda$ across the four analyses, discussed in the text.}
\label{fig:corner_case2}
\end{figure}

Figure~\ref{fig:corner_case2} shows the $\beta$--$\lambda$ degeneracy behind the $\delta_{\rm eff}$ behavior described above, together with the same $H_0$--$\Omega_{m0}$ anticorrelation already seen in Case~I.

\subsection{Case III: CPL Dark Energy with a Constant Coupling}\label{sec:results_case3}

Case~III keeps the constant coupling $\delta$ of Case~I but replaces the constant $\omega$ with the CPL form $w(a)=w_0+w_a(1-a)$. As in Case~I, and unlike Case~II, $\delta$ is sampled directly, so its dataset-dependence can be read without separating it from a prior-volume effect. Table~\ref{tab:case3_results} lists the constraints on $\{H_0,\Omega_{m0},w_0,w_a,\delta,m_\nu\}$; here $m_\nu$ is itself a sampled parameter rather than derived from $\Omega_{\nu0}$.

\begin{table}[H]
\centering
\begin{tabular}{lcccc}
\hline\hline
Parameter & CMB+DESI DR2 & +Pantheon$^{+}$ & +DES-Dovekie & +Union3 \\
\hline
$H_0\ [\mathrm{km\,s^{-1}\,Mpc^{-1}}]$ & $68.36^{+0.68}_{-0.73}$ & $68.18^{+0.61}_{-0.85}$ & $69.52^{+0.89}_{-0.84}$ & $67.61^{+0.68}_{-0.71}$ \\
$\Omega_{m0}$                          & $0.3021^{+0.0073}_{-0.0066}$ & $0.3075^{+0.0061}_{-0.0057}$ & $0.2904^{+0.0048}_{-0.0030}$ & $0.3098^{+0.0098}_{-0.0097}$ \\
$w_0$                                  & $-1.012^{+0.042}_{-0.040}$ & $-0.980^{+0.039}_{-0.068}$ & $-0.964^{+0.033}_{-0.068}$ & $-0.987^{+0.041}_{-0.066}$ \\
$w_a$                                  & $-1.388^{+0.412}_{-0.364}$ & $-0.419^{+0.25}_{-0.335}$ & $-1.113^{+0.20}_{-0.213}$ & $-0.639^{+0.588}_{-0.374}$ \\
$\delta$                               & $-0.196^{+0.140}_{-0.275}$ & $-0.177^{+0.165}_{-0.293}$ & $-0.292^{+0.091}_{-0.096}$ & $-0.108^{+0.242}_{-0.188}$ \\
$m_\nu\ [\mathrm{eV}]$ (95\% CL)       & $<0.3$ & $<0.27$ & $<0.25$ & $<0.29$ \\
\hline\hline
\end{tabular}
\caption{Marginalized posterior constraints (median with $68\%$ credible intervals) on the Case~III (CPL dark energy, constant $\delta$) parameters, for the four dataset combinations analyzed in this work. The neutrino mass $m_\nu$ is quoted as a $95\%$ CL upper limit.}
\label{tab:case3_results}
\end{table}

\begin{figure}[H]
\centering
\includegraphics[width=0.85\textwidth]{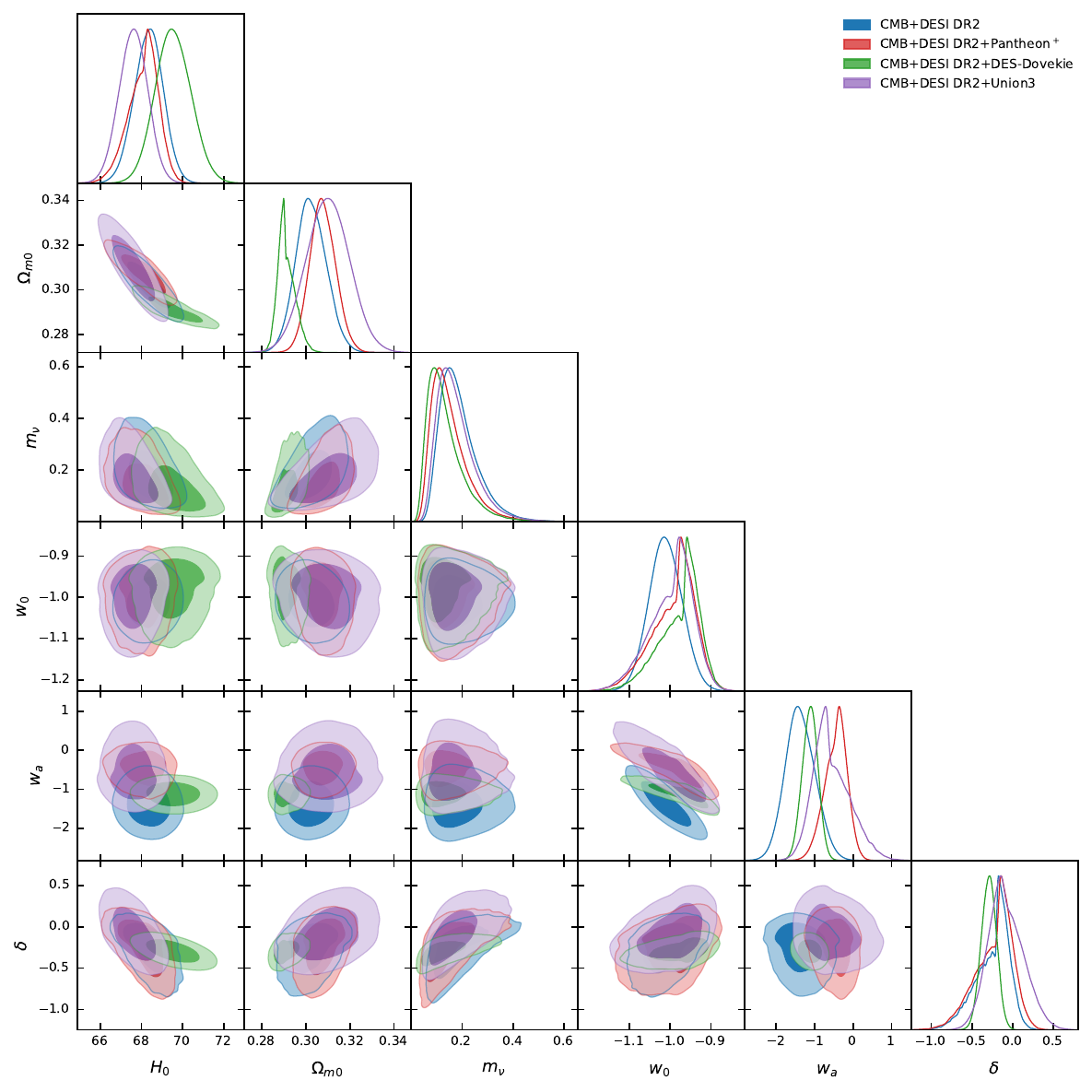}
\caption{Marginalized one-dimensional posterior distributions (diagonal panels) and two-dimensional $68\%$ and $95\%$ credible regions (off-diagonal panels) for the Case~III parameters $\{H_0,\,\Omega_{m0},\,m_\nu,\,w_0,\,w_a,\,\delta\}$, for the four dataset combinations of Table~\ref{tab:case3_results}.}
\label{fig:corner_case3}
\end{figure}

Ordering the four combinations by coupling strength, $\delta=-0.292$ (DES-Dovekie) $\to -0.196$ (CMB+DESI~DR2) $\to -0.177$ (Pantheon$^{+}$) $\to -0.108$ (Union3), $H_0$ falls and $\Omega_{m0}$ rises to match -- the same $H_0$--$\Omega_{m0}$--$\delta$ pattern visible in Fig.~\ref{fig:corner_case3}. A stronger coupling favors a higher $H_0$ and a lower $\Omega_{m0}$, with DES-Dovekie and Union3 at opposite ends of the sequence.

The neutrino mass does not follow this ordering. Its four $95\%$ CL upper limits, $m_\nu<0.3\,\mathrm{eV}$ (CMB+DESI~DR2), $<0.27\,\mathrm{eV}$ (Pantheon$^{+}$), $<0.25\,\mathrm{eV}$ (DES-Dovekie), and $<0.29\,\mathrm{eV}$ (Union3), are close to one another, so $m_\nu$ is largely insensitive to the choice of supernova sample. These bounds are markedly looser than the corresponding limits in Case~I and Case~II, consistent with $m_\nu$ being sampled directly here rather than being derived from other parameters.

$w_0$ stays close to $-1$ in all four combinations, $[-1.012,-0.964]$, unlike the phantom value found for CMB+DESI~DR2 alone in Case~I; with a time-evolution term available, the fit no longer needs to push $w_0$ away from $-1$. That term, $w_a$, is instead the least stable parameter in the table, from $-1.388^{+0.412}_{-0.364}$ (CMB+DESI~DR2) to $-0.419^{+0.25}_{-0.335}$ (Pantheon$^{+}$), without following the ordering set by $\delta$. This suggests $w_a$ absorbs supernova-specific systematics that are largely separate from the coupling degeneracy, much as $m_\nu$ does.

The total $\chi^2$ scales with the size of each supernova sample and shows no sign of tension once this is accounted for. One point is worth checking before quoting these numbers further: for CMB+DESI~DR2 alone, $\chi^2_{\rm tot}=16.43$ exceeds $\chi^2_{\rm BAO}=11.08$ by $5.35$, the presumed CMB contribution; for the three combinations with a supernova sample, however, $\chi^2_{\rm tot}$ equals $\chi^2_{\rm BAO}+\chi^2_{\rm SN}$ exactly, with no separate CMB term. It would help to confirm how this term is counted before using $\chi^2_{\rm tot}$ for model comparison. The posterior range for $\chi^2_{\rm tot}$ is also noticeably wider on one side once a supernova sample is included -- e.g. $[1772.31,30307.37]$ against a central value of $1761.19$ for DES-Dovekie -- which may reflect a small number of chain samples in poorly-constrained regions of $(w_0,w_a,\delta,m_\nu)$ space rather than a second solution; checking the trace plots and the Gelman--Rubin statistic for these three runs would settle the point.

Across the four combinations, $H_0$ in Case~III ranges from $67.6$ to $69.5\,\mathrm{km\,s^{-1}\,Mpc^{-1}}$, showing the same dataset-dependence already seen in Case~I. Here, though, this comes together with a neutrino mass upper limit around $0.3\,\mathrm{eV}$ that barely changes with the supernova sample, so the partial shift in $H_0$ and the looser mass bound should be read as parts of the same fit rather than as separate results.

\subsection{Comparison with \texorpdfstring{$\Lambda$CDM}{LCDM}}\label{sec:lcdm_comparison}

For reference, Table~\ref{tab:lcdm_comparison} lists the flat $\Lambda$CDM constraints on $H_0$ and $\Omega_{m0}$, together with the Bayesian log-evidence $\ln Z$ for $\Lambda$CDM and for the three MaVaN cases discussed above, all obtained from the same four dataset combinations.

\begin{table}[H]
\centering
\begin{tabular}{llcccc}
\hline\hline
Model & Quantity & CMB+DESI DR2 & +Pantheon$^{+}$ & +DES-Dovekie & +Union3 \\
\hline
$\Lambda$CDM & $H_0\ [\mathrm{km\,s^{-1}\,Mpc^{-1}}]$ & $68.22\pm0.30$ & $68.31\pm0.28$ & $70.09\pm0.14$ & $68.42\pm0.29$ \\
             & $\Omega_{m0}$                          & $0.2991\pm0.0039$ & $0.2982\pm0.0038$ & $0.2819\pm0.0019$ & $0.3011\pm0.0039$ \\
             & $\ln Z$                                & $14.82$ & $19.49$ & $15.75$ & $14.83$ \\
\hline
Case~I       & $\ln Z$                                & $18.15$ & $17.03$ & $18.12$ & $18.61$ \\
Case~II      & $\ln Z$                                & $18.71$ & $24.85$ & $20.44$ & $18.91$ \\
Case~III     & $\ln Z$                                & $20.63$ & $20.46$ & $24.01$ & $18.81$ \\
\hline\hline
\end{tabular}
\caption{Flat $\Lambda$CDM constraints on $H_0$ and $\Omega_{m0}$, and the Bayesian log-evidence $\ln Z$ for $\Lambda$CDM and for Cases~I--III, for the four dataset combinations used throughout this work.}
\label{tab:lcdm_comparison}
\end{table}

The $\Lambda$CDM fit already shows part of the pattern seen throughout this section: $H_0$ and $\Omega_{m0}$ move in opposite directions once a supernova sample is added, and DES-Dovekie again stands apart, with the highest $H_0$ ($70.09\pm0.14\,\mathrm{km\,s^{-1}\,Mpc^{-1}}$), the lowest $\Omega_{m0}$ ($0.2819\pm0.0019$), and by far the tightest uncertainties of the four combinations. This suggests that some of the dataset-dependence seen for the coupled models is already present in the data themselves, rather than introduced by the coupling.

The log-evidence values let us compare the four models directly, through $\Delta\ln Z = \ln Z_{\rm model} - \ln Z_{\Lambda\rm CDM}$. Using the qualitative bins commonly adopted for this kind of comparison ($|\Delta\ln Z|<1$: inconclusive; $1$--$2.5$: weak; $2.5$--$5$: moderate; above $5$: strong), the picture is mixed rather than one-sided. Case~III shows the most consistent preference over $\Lambda$CDM, moderate to strong in three of the four combinations and reaching $\Delta\ln Z \approx 8.3$ with DES-Dovekie, though this drops to essentially no preference ($\Delta\ln Z \approx 1.0$) with Pantheon$^{+}$. Case~II gives moderate to strong support in every combination, with its largest margin ($\Delta\ln Z \approx 5.4$) also coming from Pantheon$^{+}$. Case~I is the least consistent of the three: moderately preferred with CMB+DESI~DR2 alone and with Union3, only weakly so with DES-Dovekie, and mildly disfavored relative to $\Lambda$CDM once Pantheon$^{+}$ is used ($\Delta\ln Z \approx -2.5$).

Taken together, these numbers suggest that a coupling between neutrinos and dark energy is generally favored by DESI~DR2 BAO and the CMB prior, but the strength of that preference -- and, for Case~I, even its direction -- depends on which supernova compilation is used alongside them. None of the three cases is preferred over $\Lambda$CDM in every combination, so at this stage the evidence for the coupling is best read as encouraging rather than conclusive.

\section{Conclusions}\label{sec_5}

We have investigated three related models of Mass-Varying Neutrino (MaVaN) cosmology, distinguished by the way the interaction between the neutrino and dark-energy sectors is parameterized: constant coupling with a constant dark-energy equation-of-state (Case I); exponential neutrino mass coupled with an exponential quintessence potential under the adiabatic minimum-tracking assumption (Case II); and constant coupling with a CPL dark-energy equation-of-state (Case III). We compared these models against DESI DR2 BAO measurements and compressed CMB likelihoods, while also utilizing Pantheon$^{+}$, DES-SN5YR, and Union3 supernova data. This combination allows us to examine both the dependence of inferred cosmological parameters on the supernova datasets and the sensitivity of coupling constraints to the model parameterization.

A common feature across all three cases is that combining different supernova samples with DESI DR2 and CMB priors results in correlated variations in $H_0$ and $\Omega_{m0}$. The results also demonstrate that the inferred dark-energy and coupling parameters differ among the various MaVaN models. Specifically, the phantom-like value of the constant equation-of-state parameter obtained from CMB+DESI DR2 in Case-I is not observed in Case-II or Case-III; this indicates that this feature depends on the assumed parameterization rather than being a model-independent result derived from the data. In Case-II, the effective coupling $\delta_{\rm eff}=3\beta/(\beta+\lambda)$ can vary significantly across different analyses, illustrating the sensitivity of the inferred interaction to the priors adopted for mass and potential parameters. Similarly, neutrino-mass constraints vary across cases and should therefore be interpreted as model-dependent limits rather than model-independent bounds on the absolute neutrino mass scale.

Bayesian evidence provides a complementary assessment of the models. Although some dataset combinations show positive evidence differences compared to flat $\Lambda$CDM, none of the MaVaN models considered here exhibit a consistent preference across all four observational combinations. Notably, the substantial evidence difference found for Case-III with DES-SN5YR is driven by a supernova combination that also induces a distinct shift in the corresponding $\Lambda$CDM parameters. Furthermore, the varying prior ranges adopted for $\beta$ and $\lambda$ in the Case-II analysis introduce an additional prior-volume dependence into the evidence calculations. Consequently, the current results should be viewed as evidence that MaVaN interactions are observationally viable, rather than as a statistically robust detection of neutrino-dark-energy coupling.

A key feature of this analysis is the direct comparison of multiple MaVaN models within a unified observational framework, alongside a systematic examination of their dependence on datasets and priors. However, certain limitations should be noted. Case-II relies on the adiabatic minimum-tracking approximation, the validity of which requires verification within the relevant posterior parameter space. This analysis is restricted to the background level and does not account for the perturbation dynamics of the interacting neutrino-scalar system, neutrino clustering, or potential small-scale instabilities. Furthermore, the compressed CMB likelihood does not capture the full information contained in the CMB power spectra. The use of varying prior choices in some analyses also limits the extent to which the associated Bayesian evidence can be interpreted as a purely model-based comparison. These limitations motivate several natural extensions of the current work, such as adopting a common and clearly defined prior volume for all model variants, considering full scalar-field dynamics without the adiabatic approximation, incorporating linear perturbations and neutrino clustering, and utilizing the full CMB likelihood instead of a compressed representation. Applying these extensions to upcoming DESI data releases and updated supernova compilations will help determine whether evidence for neutrino-dark-energy coupling persists independently of model assumptions, prior choices, and the specific low-redshift distance datasets employed.

\begin{acknowledgments}
S. D. Pathak acknowledges the Inter-University Centre for Astronomy and Astrophysics (IUCAA), Pune, for the Associate Programme under which this work was carried out.
\end{acknowledgments}

\bibliographystyle{elsarticle-num}
\bibliography{mybib.bib}

\end{document}